%% file: samplepaper.tex
\pdfoutput=1
\documentclass[runningheads]{llncs}
\usepackage{graphicx}
\usepackage{multirow}
\usepackage{multicol}
\usepackage[ruled]{algorithm2e}
\usepackage{subfigure}
\usepackage{array}
\usepackage{float}
\usepackage{algorithmic}
\usepackage{cite}
\usepackage{amsmath,amssymb,amsfonts}
\usepackage{textcomp}
\usepackage{xcolor}
\usepackage{hyperref}
\usepackage[hyphenbreaks]{breakurl}
\begin{document}
\title{DeepHunter: A Graph Neural Network Based Approach for Robust Cyber Threat Hunting}
\titlerunning{DeepHunter: A GNN-based Approach for Robust Threat Hunting}
\author{Renzheng Wei\inst{1,2} \and
Lijun Cai\inst{1} \and
Aimin Yu\inst{1} \and
Dan Meng\inst{1}}

%\author{Renzheng Wei\inst{1}\orcidID{0000-1111-2222-3333} \and
%Lijun Cai\inst{2,3}\orcidID{1111-2222-3333-4444} \and
%Aimin Yu\inst{3}\orcidID{2222--3333-4444-5555} \and
%Dan Meng\inst{3}\orcidID{2222--3333-4444-5555}}
%
\authorrunning{R. Wei et al.}
%\institute{Institute of Information Engineering,  Chinese Academy of Sciences, Beijing, China \and
%Springer Heidelberg, Tiergartenstr. 17, 69121 Heidelberg, Germany

\institute{Institute of Information Engineering,  Chinese Academy of Sciences, Beijing, China \and
School of Cyber Security, University of Chinese Academy of Sciences, Beijing, China\\
\email{\{weirenzheng,cailijun,yuaimin, mengdan\}@iie.ac.cn}}

\maketitle              % typeset the header of the contribution
\begin{abstract}
Cyber Threat hunting is a proactive search for known attack behaviors in the organizational information system. It is an important component to mitigate advanced persistent threats (APTs). However, the attack behaviors recorded in provenance data may not be completely consistent with the known attack behaviors. In this paper, we propose DeepHunter, a graph neural network (GNN) based graph pattern matching approach that can match provenance data against known attack behaviors in a robust way. Specifically, we design a graph neural network architecture with two novel networks: \textit{attribute embedding networks} that could incorporate Indicators of Compromise (IOCs) information, and \textit{graph embedding networks} that could capture the relationships between IOCs. To evaluate DeepHunter, we choose five real and synthetic APT attack scenarios. Results show that DeepHunter can hunt all attack behaviors, and the accuracy and robustness of DeepHunter outperform the state-of-the-art method, Poirot.
%150-250个词
\end{abstract}
\input{ccs-body}
\bibliographystyle{splncs04}
\bibliography{reference}
%Author, F.: Article title. Journal \textbf{2}(5), 99--110 (2016)
%Author, F., Author, S.: Title of a proceedings paper. In: Editor,
%F., Editor, S. (eds.) CONFERENCE 2016, LNCS, vol. 9999, pp. 1--13.
%Springer, Heidelberg (2016). \doi{10.10007/1234567890}
%Author, F., Author, S., Author, T.: Book title. 2nd edn. Publisher,
%Author, A.-B.: Contribution title. In: 9th International Proceedings
%on Proceedings, pp. 1--2. Publisher, Location (2010)
%LNCS Homepage, \url{http://www.springer.com/lncs}. Last accessed 4
%\end{thebibliography}
\end{document}

%% file: ccs-body.tex
\section{Introduction}\label{introduction}
Threat hunting is a proactive search for intruders who are lurking undetected in the organizational information system. A typical task for a threat hunter is to match system events against known adversarial behavior gained from CTI (Cyber Threat Intelligence). Threat hunting is increasingly becoming an important component to mitigate the Advanced Persistent Threats (APTs), as large enterprises or organizations seek to stay ahead of the latest cyber threats and rapidly respond to any potential attacks.  %With the knowledge from CTI, the threat hunter can execute deep analysis and discover known attacks before achieving their goals.% on the events produced by internal monitoring systems, e.g. SIEM, IPS or IDS/NIDS/HIDS. %For example, the threat hunter compares DNS logs (i.e. domains and IPs) with IOCs to look for potentially malicious activities, e.g. IP addresses and open ports used by attackers.       For example, given IOCs, the threat hunter can search for potentially attack activities (e.g. IP addresses and open ports used by attackers) from system logs.                 , e.g. malicious files, process names, domain names, etc.   %, and TTPs(adversarial Tactics, Techniques, and Procedures)     %  备选  With the help of CTI, the threat hunter can execute deep analysis, such as forensic analysis, attack tracing, and attack scenario reconstruction.

Existing threat hunting tools (e.g., Endpoint Detection and Response tools, namely EDR) rely on matching low-level Indicators of Compromise (IOCs) or TTP rules (i.e., adversarial Tactics, Techniques, and Procedures). However, simple rules matching methods are prone to high volumes of false alarms, which leads to the ``threat alert fatigue" problem. To overcome this problem, recent works~\cite{homayoun2017know,milajerdi2019poirot,hassan2020tactical} start to focus on the relationship between IOCs or the correlation among threat alerts. One approach~\cite{homayoun2017know} to hunt the ransomware takes advantage of the sequential relationship among IOCs, but the mined sequential patterns typically can not capture long-term attack behaviors.

Recent research suggests that the $provenance$ $graph$ can incorporate the long-term historical context and facilitate  threat investigation. Based on the provenance graph, many works~\cite{hassan2020tactical,milajerdi2019poirot} have made advancements to improve the performance of threat hunting. For example, RapSheet~\cite{hassan2020tactical} leverages dependency relations in the provenance graph to correlate the threat alerts generated by EDR tools, then drops the alerts that do not conform to the APT ``kill chain". Poirot~\cite{milajerdi2019poirot} improves the accuracy of threat hunting by designing a graph pattern matching algorithm to search the provenance graph for the $query$ $graph$ that represents the known attack behavior.

Although the provenance graph can greatly facilitate threat hunting tasks, there still exist several limitations in the existing approaches:

\begin{itemize}
% the real attack recorded in the provenance graph is not completely consistent with the known attack described by CTI. For example, the important path might be missing in the provenance graph (We present this scenario in Section~\ref{bme:motivatingexample}). The inconsistency would impair the performance and weaken the robustness of graph traversal-based approaches, i.e. Poirot~\cite{milajerdi2019poirot}. We will demonstrate that using the APT scenario in Section~\ref{eva:robustness}. %Many attacks could result in the incomplete provenance graph, even if the provenance graph is constructed by a whole-system provenance tracking system. (One obvious example is the side channel attacks.) Besides, some provenance tracking systems running on the application layer could also miss events that are critical for threat hunting. The incomplete provenance graph would impair the performance and weaken the robustness of graph traversal-based approaches. We will demonstrate that using a real APT example in Section 2.3 and Section 5.3.

     \item {\bf Expert knowledge needed.} Existing threat hunting tools or methods need analysts with expert knowledge on known attacks and target systems (e.g., Windows, Linux, macOS, etc.). For example, one needs to estimate the number of entry points of APT attacks when setting Poirot's threshold.   %(i.e. in Poirot~\cite{milajerdi2019poirot} and Holmes~\cite{milajerdi2019holmes}), analysts need to have expert knowledge on the related attacks; To select the appropriate parameters for model, such as the threshold in Poirot, one needs to estimate possible entry points of an APT attack.

     \item {\bf Efficiency.} The size of the provenance graph is very large because of the presence of long-term attacks. So the provenance graph-based approaches (i.e., graph matching/searching algorithms) must be efficient.

    \item {\bf Lack of robustness (most important).} In practice, real attack activities recorded in provenance data are not completely consistent with the known attack behaviors due to auditing/monitoring systems, attack mutations, and random noise. For example, one or more attack steps in CTIs might disappear in the provenance graph. This sort of inconsistency weakens the ability of provenance graph-based methods~\cite{milajerdi2019holmes,milajerdi2019poirot,hassan2020tactical} to correlate threat alerts. Even worse, the attack provenance graphs might be disconnected, which will bring errors into path-based approaches, i.e., Poirot
    ~\cite{milajerdi2019poirot}. We will detail this scenario in section~\ref{bme:motivatingexample}.
\end{itemize}

In recent years, graph neural networks (GNNs) have shown great success in handling graph data. Inspired by that, our idea is to view the threat hunting task as a graph pattern matching problem and leverage the powerful GNN model to estimate the matching score between the provenance graph and the given query graph. The graph neural networks have several advantages on the graph pattern matching problem: (1) The graph neural networks naturally excel at efficiency, since modern GPUs can largely accelerate matrix computations by parallel processing. (2) No additional expert knowledge about attacks and target systems is needed, as the graph neural network is trained in an end-to-end manner. What we need is to learn a GNN-based graph pattern matching model that could extract robust graph patterns that are resistant to the inconsistency mentioned earlier. Basically, if both the node attributes (i.e., IOC information) and the graph structures (i.e., dependency relations between IOCs) in the query graph are largely matched in the provenance graphs, the model should output a high matching score and raise alarms. 

Unfortunately, there is no off-the-shelf GNN-based architecture that can be simply applied to solve our problem due to two reasons. First, indicators are the entity with multiple attributes (i.e., file names, IP addresses, ports, process names, etc.). Different attributes may have different importance to the graph pattern matching task. Second, the two input graphs for graph pattern matching have different characteristics: The query graph is small and noise-free; The provenance graph is bigger and contains redundant nodes, as the provenance graph represents low-level system events.

To solve these problems in threat hunting, we propose two novel graph neural network structures: the \textit{attribute embedding network} and the \textit{graph embedding networks}. The attribute embedding network encodes attributes into vectors. In particular, we add the attention mechanism to the attribute embedding network. So it could assign higher weights to those attributes that are important to the graph matching task. The graph embedding networks are used for representing graph structures. To better represent distinct input graphs, we employ two different graph embedding networks to encode them, respectively. Specifically, we design one graph embedding network to represent the provenance graph and adopt GCN~\cite{kipf2016semi} to represent the query graph. At last, we utilize a powerful relation learning network (i.e., NTN~\cite{socher2013reasoning}), instead of the traditional Siamese network, to learn a metric for computing the matching score. With this new design practice, we could build the GNN model for graph pattern matching, which is robust against different degrees of inconsistency between the query graph and the provenance graph in threat hunting.

We implemented our proposed technique as \textbf{DeepHunter}, a GNN-based graph pattern matching model for threat hunting. To evaluate the accuracy and robustness of DeepHunter, we choose 5 APT attack scenarios with different degrees of inconsistency. 
%Among these scenarios, a small number of nodes and edges in the attack provenance graphs would be deleted, modified, or added. 
Particularly, one of these scenarios (Q5) contains disconnected attack provenance graphs. Experimental results show that DeepHunter can identify all of the attack behaviors in 5 APT scenarios, and it is resistant to various degrees of inconsistency and the disconnected attack provenance graphs. The robustness of DeepHunter outperforms the state-of-the-art APT threat hunting method, Poirot. Moreover, DeepHunter could find attacks that Poirot can not identify under the specific complex attack scenario, Q5+ETW. We also compare DeepHunter with other graph matching approaches, including a non-learning approach and GNN-based approaches. Results show that the performance of DeepHunter is superior to these methods.
%We measure the degree of inconsistency between the query graph and the provenance graph from three aspects: $missing$ $nodes$, $missing$ $paths$, and the graph edit distance (GED).

In summary, this paper makes the following contributions:

\begin{itemize}
	\item We propose DeepHunter, which is a GNN-based graph pattern matching approach for cyber threat hunting. DeepHunter can tolerate the inconsistency between the real attack behaviors recorded in provenance data and the known attack behaviors to some extent. %that graph neural networks are a viable approach to cyber threat hunting. More importantly, we could design a graph neural network that is robust against the inconsistency between real attack behaviors recorded by provenance data and known attack behaviors to some extent.
%    \item We discover that graph neural networks are a viable approach to cyber threat hunting. More importantly, we could design a graph neural network that is robust against the inconsistency between real attack behaviors recorded by provenance data and known attack behaviors to some extent.

    \item We design a graph neural network architecture with two novel networks: \textit{attribute embedding networks} and \textit{graph embedding networks}. These two networks could capture complex graph patterns, including IOC information and the relationships between IOCs.%We design two novel graph neural networks, i.e., attribute embedding network and graph embedding network, to incorporate Indicators of Compromise (IOCs) information and the relationships between IOCs. propose a novel graph neural network architecture which contains the attribute embedding network and the graph embedding network. We design two novel neural networks, i.e., attribute embedding network and which propose a novel graph neural network architecture which contains the attribute embedding network, the graph embedding networks, and a relation learning network. %This proposed GNN model takes advantage of valuable IOC information and could encode two different types of input graphs (i.e. the query graph and the provenance graph).

    % \item We propose a new graph neural network architecture which contains  the attribute embedding network, the graph embedding networks, and a relation learning model. Leveraging the proposed model, even if there is a certain degree of deviation between the actual attack behavior and the known attack behaviors, the threat hunter can identify them accurately.

     \item We choose 5 APT attack scenarios with different degrees of inconsistency between the provenance graph and the query graph, including 3 real-life APT scenarios and 2 synthetic APT scenarios, to evaluate our approach. %Among these scenarios, there are different degrees of inconsistency between the provenance graph and the query graph and disconnected attack provenance graphs.

     \item Our evaluation illustrates that DeepHunter outperforms the state-of-the-art APT threat hunting approach (i.e., Poirot) in accuracy and robustness. Meanwhile, DeepHunter, as a graph pattern matching model,  is superior to other graph matching methods (i.e., non-learning-based and GNN-based) in the threat hunting task. %Compared with other graph matching methods (i.e., non-learning-based and GNN-based), DeepHunter can achieve higher performance.
\end{itemize}

\input{relatedwork}

\input{bme-2}
\input{overview-2.5}

\input{gnn-3}
\input{implement-4}

\input{evaluation-5}

\vspace{-0.5in}
\section{Conclusions}
We propose DeepHunter, a GNN-based graph pattern matching approach for cyber threat hunting. More importantly, DeepHunter is robust against the inconsistency between real attack behaviors recorded by provenance data and known attack behaviors to some extent. Our extensive evaluations show that DeepHunter can tolerate various scenarios with different inconsistency scores, including disconnected attack provenance graphs. In our synthetic APT attack scenario, DeepHunter is superior to the state-of-the-art APT threat hunting approach Poirot. Our research showcased a successful application of the graph neural network on the threat hunting task.

%% file: relatedwork.tex
\section{Related Work}
%We survey related work in threat hunting approaches, provenance graph analysis, and deep graph matching approaches.

\subsection{Threat Hunting Approaches}
In this work, we mainly focus on the threat hunting methods. Poirot~\cite{milajerdi2019poirot} is a related work to DeepHunter. We will introduce and compare it with DeepHunter in the evaluation (section~\ref{eva:robustness}). RapSheet~\cite{hassan2020tactical} is an approach that could improve EDR's threat hunting ability using the provenance graph analysis. But RapSheet~\cite{hassan2020tactical} requires complete paths remained in the provenance graph to correlate alerts. Obviously, the disconnected attack provenance graphs will undermine the performance of RapSheet.

For APT detection and investigation, both Holmes~\cite{milajerdi2019holmes} and NoDoze~\cite{hassan2019nodoze} correlate alerts using the provenance graphs. To hunt stealthy malware, ProvDetector~\cite{wang2020you} proposes a graph representation learning approach to model process' normal behavior in provenance graphs. However, these methods assume an accurate normal behavior database for reducing false alarms. We know that the normal behavior model may create a risk of the poisoning attack due to concept drift as benign usage changes. Additionally, all of these methods are path-based approaches. So their robustness could be influenced by the disconnected provenance graphs.

Some methods use IOCs or threat alerts as a clue to identify attack behaviors (i.e., zero-day attack~\cite{sun2018using} and C\&C~\cite{oprea2015detection}). However, these methods overlook the relationship between indicators or alerts. So it could bring high false positives.

\subsection{Provenance Graph Analysis}
Provenance graph analysis is widely applied to the APT attack detection~\cite{xiong2020conan}, forensic analysis~\cite{hossaincombating}, and attack scenario reconstruction~\cite{pei2016hercule,hossain2017sleuth}, etc. Recent works~\cite{hassan2020omegalog,milajerdi2019holmes} seek to bridge the semantic gap between low-level system events and high-level behaviors. Many recent works (i.e., Morse~\cite{hossaincombating}, BEEP~\cite{lee2013high}, MPI~\cite{ma2017mpi}, and OmegaLog~\cite{hassan2020omegalog}, etc.) are proposed to address the dependency explosion problem in provenance graphs. StreamSpot~\cite{manzoor2016fast} views the provenance graph as a temporal graph with typed nodes and edges, then proposes a graph sketching algorithm for anomaly detection.

\subsection{Graph Matching Approaches}
%We have compared one graph kernel based and three GNN based graph matching models earlier in this paper.
Graph pattern matching and graph similarity computation have been studied for many real applications, such as the binary function similarity search~\cite{pmlr-v97-li19d} and the hardware security~\cite{fyrbiak2019graph}. In the past few decades, many graph matching metrics were defined, e.g., graph edit distance, graph isomorphism, etc. In this paper, we evaluate the non-learning based WL kernel as a graph matching method. Recently, many graph neural networks have been proposed for graph pattern matching. We compare three of them in the evaluation. %Although there still exist other graph matching models (e.g., GraphSim~\cite{bai2020learning}), no one is suitable for the threat hunting task.

%% file: bme-2.tex
\vspace{0.5in}
\section{Background and Motivation}
In this section, we first introduce the background knowledge of the provenance graph and the query graph. (Section~\ref{bme:backgroud}). Then, we briefly illustrate several common motivating situations where the threat hunting approach calls for high robustness (Section~\ref{bme:motivatingsituations}). Finally, taking an APT attack scenario with disconnected attack graphs as an example, we illustrate how this scenario affects existing methods and explain why DeepHunter can resist this situation (Section~\ref{bme:motivatingexample}). %At last, we give the problem statement and scope (Section~\ref{bme:problemstatement}).

\subsection{Background}\label{bme:backgroud}

\subsubsection{Provenance Graph.}\label{bme:provenanceGraph}
Provenance graph is generally a directed acyclic graph (DAG)~\cite{gibson2009application}, where the nodes represent system entities, and the edges represent the dependency relation between these entities. There are two types of nodes in provenance graphs: subjects, which represent processes, and objects, which represent other system entities such as files, Windows registry, and network sockets, etc. Subject node's attributes include process name, command line arguments. Object node's attributes include file names, IP addresses, ports, etc. Table~\ref{Tabel1} shows the nodes and edges we consider in this work. Provenance graph can represent dependencies between system events. %And it is commonly used to track attack traces or assess the ramification of attacks. 

\begin{table}[]
\centering
\caption{A summarization of nodes and edges in provenance graphs and query graphs.}
\renewcommand{\arraystretch}{1.1}
\begin{tabular}{p{2.5cm}<{\centering}|p{2.2cm}<{\centering}|p{3.5cm}<{\centering}|p{2.2cm}<{\centering}}   % p{x com} x��ʾ����xcm�Զ�����
%\begin{tabular}{c|c|p{2.8cm}|c}   % p{x com} x��ʾ����xcm�Զ�����
\hline
\textbf{Subject   type}  & \textbf{Object   type}        & \textbf{Attributes}                             & \textbf{Relations} \\ \hline \hline
\multirow{4}{*}{Process} & Process                       & Name, Augments                                  & Fork/Clone                 \\ \cline{2-4}
                         & File                          & File   name                                     & Read, Write                 \\ \cline{2-4}
                         & Socket                        & Src/dst IP, Src/dst port                        & Recv, Send                  \\ \cline{2-4}
                         & Registry                      & Key   name                                      & Write                      \\ \hline \hline
\end{tabular}

\label{Tabel1}
\end{table}

\subsubsection{Query Graph.}
The query graph $G_{q}$ in our work can be constructed by manually or automatically~\cite{husari2017ttpdrill,liao2016acing,zhu2018chainsmith} extracting IOCs together with the relationships among them from CTIs (including human-written reports or other threat intelligence feeds with structured standard formats (e.g., STIX~\cite{stix}, OpenIOC~\cite{openioc} and MISP~\cite{misp})). Both nodes and edges of the query graph are the same as the provenance graph, as shown in Table~\ref{Tabel1}. We set the node's or edge's attributes to null if CTIs do not include the corresponding information.

\subsection{Motivating Situations}\label{bme:motivatingsituations}
To motivate our work, we introduce several common situations that could lead to inconsistency or disconnected provenance graphs, which will weaken the existing methods' threat hunting ability. Firstly, provenance systems (e.g., Spade~\cite{gehani2012spade}) that are developed for recording system events in the application layer may overlook certain attack activities. For instance,  Spade does not trace system activities until user space is started. Hence, if the attack occurred before the tracing of Spade, an incomplete attack provenance graph would be generated. Fortunately, the whole-system provenance trackers (like Hi-Fi~\cite{pohly2012hi}, LPM~\cite{bates2015trustworthy}, and CamFlow~\cite{pasquier2017practical}) can overcome this problem. Because these whole-system provenance trackers begin recording system activities in the early boot phase as the $INIT$ process starts. 

%constructs provenance graphs using audit data streams collected from the application layer. Nevertheless, in some cases Spade may not be able to record a complete attack provenance graph. For instance, since Spade does not trace behavior until user space is started, an incomplete provenance graph is generated if the attack occurred before the tracing of Spade. Fortunately, the whole-system provenance tracker can overcome this problem. The whole-system provenance trackers, like Hi-Fi~\cite{pohly2012hi}, LPM~\cite{bates2015trustworthy}, and CamFlow~\cite{pasquier2017practical}, are kernel-layer auditing frameworks that begin recording system activity in the early boot phase as the $INIT$ process starts. 

%Hence the whole-system provenance tracker can overcome the above mentioned problem.
%The provenance graph can model system behavior with different granularity.
%But Spade does not trace behavior until user space is started. So if the attack occurred before the tracing of Spade, the provenance graphs generated by it is incomplete. 

Secondly, even if the whole-system provenance system is applied, some targeted attacks could also lead to inconsistency or disconnected attack provenance graphs. Taking the microarchitectural side-channel attack as an example, there is no connection between the attacker process and the victim process in the provenance graph. Coordinated attacks could generate disconnected attack provenance graphs as well. If attackers control multiple entry points of a compromised system and coordinate to achieve an operational goal, each entry point may correspond to an isolated attack graph. 
%Secondly, even if the whole-system provenance system is applied, some targeted attacks could also lead to inconsistency or disconnected attack provenance graphs. We briefly introduce three of these attacks as follows. \textcircled{1} Experienced attackers usually compromise audit logs at the end of an advanced attack. Instead of directly wipe the associated logs~\cite{incidentResponseReport,destroyinglogs}, the attacker alternatively tampers the logs (i.e., edits existing events or deletes important attack paths)~\cite{logcleaner-github}, since the contaminated logs could confuse investigators. \textcircled{2} Some specific side-channel attacks (e.g., microarchitectural side-channels) are another factor leading to the disappearance of attack paths in the whole-system provenance graphs. \textcircled{3} $kauditd$ buffer overflows can also cause log loss if the provenance system is based on the Linux Audit Framework.

%Thirdly, coordinated attacks could generate disconnected attack provenance graphs as well. If attackers control multiple entry points of a compromised system and coordinate to achieve an operational goal, each entry point may correspond to an isolated attack graph. 

Finally, attack mutations (or inaccurate CTIs) are another reason that incurs the inconsistency or disconnected attack provenance graphs. In the next section, we present an APT scenario with attack mutations to illustrate its influence to threat investigation. %We detail this situation by presenting an APT attack scenario in the next section.% and analysis its influence to threat hunting methods.

\subsection{An APT Attack Scenario}\label{bme:motivatingexample}
\begin{figure}
\centering
\includegraphics[width=1.0\textwidth]{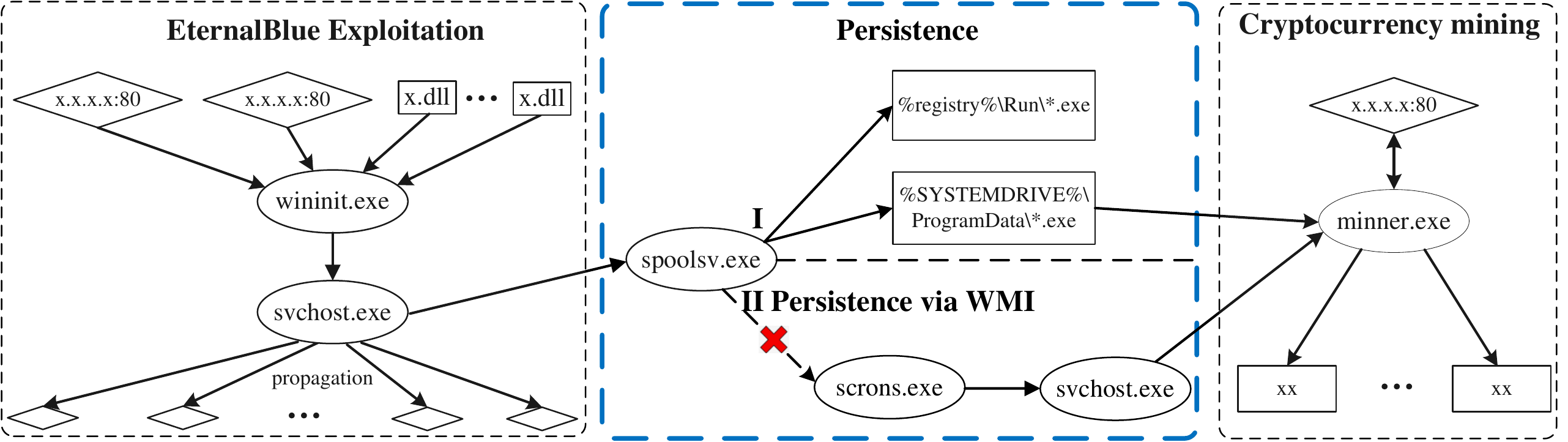}
\caption{Query graph of synthetic APT attack scenarios illustrated in section~\ref{bme:motivatingexample}.} \label{fig1}
\end{figure}

Recently, cryptocurrency mining malware is one of the most prevalent threats in the wild. Fig.~\ref{fig1} describes a typical cryptominer's progression, including the EternalBlue exploitation stage, the persistence stage, and the cryptocurrency mining stage. At its exploitation stage, $wininit$ is responsible for configuring and reconnaissance scans. $svchost$ exploits the EternalBlue vulnerability for propagation. At the persistence stage (persistence I), $spoolsv.exe$ creates an executable binary and adds its path to the ``run keys" in the Windows registry. At the last stage, the cryptominer process $minner.exe$ is started.

Now, let's consider what happens if the attacker changes the persistence techniques. 
For example, the attacker adopts an alternative persistence technique \uppercase\expandafter{\romannumeral2} (persistence via WMI in Fig.~\ref{fig1}). 
WMI is a preinstalled system tool and it can achieve fileless attacks~\cite{wmiaptreportwebpage,graeber2015abusing}.
We further assume that the running provenance system (not a whole-system provenance system) can not capture the dependency between the $spoolsv.exe$ process and the $scrons.exe$ process (which is the host process of WMI script).
Hence, the connection between the EternalBlue exploitation stage and the cryptocurrency mining stage is broken in attack provenance graphs. 

Note that the query graph used by analysts is the EternalBlue exploitation stage, the cryptocurrency mining stage, and the upper part of the persistence stage (persistence technique \uppercase\expandafter{\romannumeral1}) in Fig.~\ref{fig1}. 
So the behavior recorded in the attack provenance graph is inconsistent with the given query graph. 
Besides, the attack provenance graph is disconnected.
This situation makes both threat hunting and forensic investigation more difficult. 
In Section~\ref{eva:robustness}, we will show that this attack mutation can seriously impair the existing threat hunting approach (i.e., Poirot). 
Additionally, existing provenance graph-based threat correlation methods, like ~\cite{hassan2019nodoze,milajerdi2019holmes,hassan2020tactical}, will definitely lose the correlation between the alerts of the exploitation stage and the alerts of the cryptocurrency mining stage. 
And the path-based anomaly scores (e.g., rareness score~\cite{hassan2019nodoze,wang2020you} and threat score~\cite{hassan2020tactical}) may also be affected by disconnected attack provenance graphs.

In contrast, DeepHunter is robust against this attack mutation. Intuitively, although there exist the inconsistencies and the disconnected attack graphs in this scenario, most node attributes and the main graph structures are preserved. DeepHunter can learn robust graph patterns from training data which are resistant to the inconsistencies. We detail the design of DeepHunter in section~\ref{gnn}.

%% file: overview-2.5.tex
\section{Design Overview and Challenges}\label{approach:overview}

\subsection{Graph Pattern Matching for Cyber Threat Hunting}
We aim to determine if a provenance graph and a given query graph represent the same attack behaviors for a threat hunting task. %We know that there exists the semantic gap between low-level provenance data and high-level attack behaviors. Previous approaches to reveal concise high-level attack behaviors from provenance graphs usually adopted graph summarization~\cite{hassan2018towards} or graph sketching techniques~\cite{han2020unicorn}.%, which need domain knowledge on the attack and target systems.  %we use the graph pattern matching model $\mathcal{M}$ which could compute matching scores between them. %for investigatirefine the provenance graph by using graph summarization or graph sketching techniques, which could .
In this work, we formulate the threat hunting task as a graph pattern matching problem. Given a query graph $G_{q}$, the output is a matching score $s$ of $(G_{q}, G^{i}_{p})$, where $G^{i}_{p} \in S=\{G^{1}_{p},G^{2}_{p},\ldots,G^{N}_{p}\}$, $S$ is the set of provenance graphs. Our goal is to learn a graph matching model $\mathcal{M}$, where $\mathcal{M}(G_{p}, G_{q})=1$ indicates that the provenance graph $G_{p}$ and the query graph $G_{q}$ represent the identical behavior; otherwise, $\mathcal{M}(G_{p}, G_{q})=-1$ indicates that they are different. The graph pattern matching model $\mathcal{M}$ must meet three requirements: \textcircled{1} No expert knowledge needed; \textcircled{2} High efficiency (Graph pattern matching is NP-complete in the general case.); \textcircled{3} High robustness.

%Specifically, we hope to learn a $\mathcal{M}$, such that $\mathcal{M}(G_{p}, G_{q})=1$ indicates that the provenance graph $G_{p}$ and the query graph $G_{q}$ represent the identical behavior; otherwise, $\mathcal{M}(G_{p}, G_{q})=-1$ indicates that they are different. The graph pattern matching model $\mathcal{M}$ must meet three requirements: \textcircled{1} No expert knowledge needed; \textcircled{2} High efficiency; \textcircled{3} High robustness.

As aforementioned, using a graph neural network to extract graph patterns and further compute matching scores is particularly appealing, since it can learn a graph matching model without expert knowledge. Also, once the graph matching model is learned, the matching score can be efficiently computed, and thus we no longer rely on any expensive graph pattern matching algorithms.

\vspace{-0.001in}
\subsection{Challenges \& Solutions}

It has been demonstrated that graph neural networks can learn complex graph patterns for downstream tasks, such as binary code similarity detection~\cite{xu2017neural} and memory forensic analysis~\cite{song2018deepmem}. In this work, we need the graph neural network to extract graph patterns for matching two graphs. In particular, the graph patterns should be composed of node attributes and graph structures. We show the challenges of designing graph neural networks and our corresponding solutions as follows.

\subsubsection{Challenge 1}
How to represent node attribute information effectively? There are multiple node types in graphs, each node has many attributes, and different attributes may have different importance to the graph pattern matching task. Previous work~\cite{manzoor2016fast} considers the provenance graph as a heterogeneous graph and searches the heterogeneous graph following the meta-paths. However, constructing the meta-path needs expert knowledge on the target systems. 

We propose the attribute embedding network (detailed in section~\ref{gnn:aen}) to represent the node's attributes. We treat the node type (e.g., process, file, socket, etc.) as one of a node's attributes and employ the attention mechanism to automatically learn which attributes contribute most to the graph matching task.

%Firstly, to incorporate node attributes information, we propose the attribute embedding network (detailed in section~\ref{gnn:aen}) which could encode the node attributes into a vector. Specifically, we employ the attention mechanism to automatically learn which attributes contribute most to the graph matching task. Note that we treat the node type (i.e., process, file, socket, registry.) as one of the attributes of a node, instead of considering provenance graphs as a heterogeneous graph~\cite{manzoor2016fast} and searching the heterogeneous graph following the meta-paths. Because constructing the meta-path needs expert knowledge on the target systems. %different with existing models (e.g., ~\cite{manzoor2016fast}) that view the provenance graph as a heterogeneous graph and search the provenance graph following the meta-paths, . Unfortunately, constructing the meta-path needs expert knowledge on the target systems. Differently, we treat the node type as one of the attributes of a node, and employ the attention mechanism to automatically learn which attributes contribute most to the graph matching task. We call this model as the attribute embedding network which is illustrated in section~\ref{gnn:aen}.

\subsubsection{Challenge 2}
How to represent graph structures effectively? Previous graph pattern matching models~\cite{pmlr-v97-li19d,bai2019simgnn,wang2019heterogeneous} utilize the same neural network structure to represent both input graphs. But the characteristics of two input graphs for threat hunting are distinct, as mentioned in section~\ref{introduction}. 

We adopt two different graph neural networks: One is GCN for the query graph, and the other is specially designed to represent the provenance graph structure. We introduce them as the graph embedding networks, as detailed in section~\ref{gnn:gen}.

%% file: gnn-3.tex
\section{DeepHunter's Graph Pattern Matching Model}\label{gnn}
%\section{GRAPH NEURAL NETWORKS BASED MODEL}\label{gnn}

\subsection{Attribute Embedding Network for Encoding Node's Attributes}\label{gnn:aen}
\begin{figure*}
\centering
\includegraphics[width=1.0\textwidth]{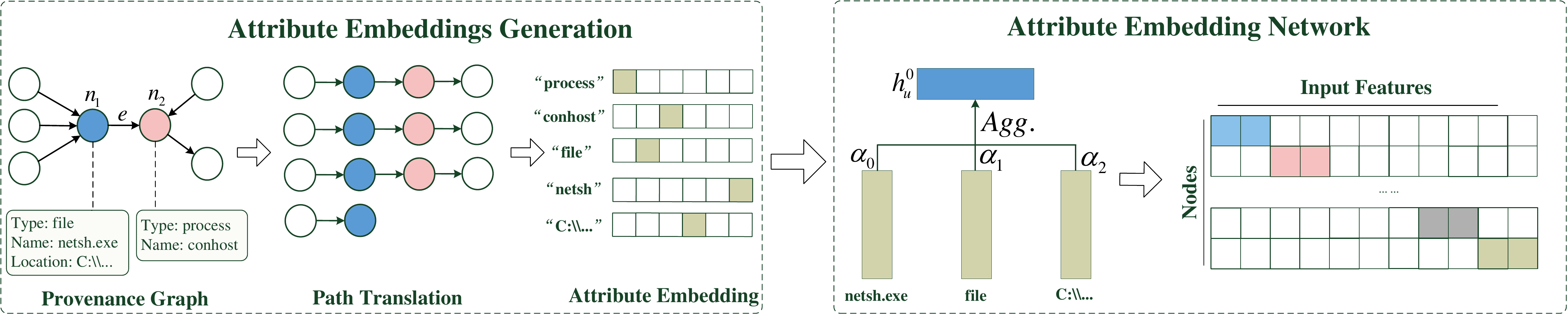}
\caption{The process of generating input features.} \label{fig5}
\end{figure*}

The goal of the attribute embedding network is to obtain the input feature $h^{0}_{u}$ for each node $u$, which incorporates $u's$ attributes information. Specifically, we first generate an embedding $v_{i}$ for each attribute $i$ of the node $u$ (as depicted on the left of Fig.~\ref{fig5}), and then compute $u's$ input feature $h^{0}_{u}$ by aggregating $u's$ attribute embeddings (as depicted on the right of Fig.~\ref{fig5}).
%The input of the attribute embedding network is raw IOC information (e.g., types, file names, etc.). 
%Many graph neural networks aimed at encoding graph structure information may use the randomly initialized vector or the one-hot encoding as the node's input feature. However, to well incorporate the node attribute information into graph embeddings, we propose to represent the node by its attribute embeddings.% which are pre-trained vectors.

%Specifically, we first obtain a vector representation for each attribute of a node. 
To obtain the attribute embedding $v_{i}$, inspired by the path embedding method of ProvDetector~\cite{wang2020you}, we view a path in the provenance graph as a sentence and then adopt an unsupervised NLP model ( word2vec~\cite{le2014distributed}). Specifically, we first translate paths in the provenance graphs into sentences which consist of attributes. For example, the colored nodes $n_1$, $n_2$, and the edge $e$ between them in the provenance graph of Fig.~\ref{fig5} can be translated into a sentence as follows: Process $conhost$ reads file $netsh.exe$ in $C:\backslash\backslash Windows \backslash \backslash System32$. Then we feed the sentences into a word2vec model to learn the vector representation $v_{i}$ for each attribute $i$. %in  for example to illustrate how to translate a path into a sentence. The text in a solid rectangular box is the attributes of a node. So the translated sentence is as follows: 

%HERE!!!!!!!!!!!!!
%Its attributes contain the type: $file$, the name: $netsh.exe$, and the file location. Now, we translate the rectangle node, the circular node and the edge $e_1$ between them in Fig.~\ref{fig5} as a sentence: Process $conhost$ read file $netsh.exe$ in $C:\backslash \backslash Windows \backslash \backslash System32$. Then we feed the sentences into the word2vec model to learn the vector representation of each attribute. We feed translated sentences into Skip-gram model and get the embedding of each attribute.

We represent a node $u's$ input feature as the aggregation of its attribute embeddings $v_{i}$. Common aggregation functions include $sum$ and $average$. However, for the graph pattern matching task, the importance of each node attribute may be different. Hence, we use the attention mechanism to learn the weight for each attribute of a node. Specifically, we compute node $u's$ input feature $h^{0}_{u}$ by

\begin{equation}
h^{0}_{u}=\sum\nolimits_{i \in A_{u}} \alpha_{i} v_{i}, \label{equ:attribute embed}
\end{equation}

\noindent
where $A_{u}$ is the attribute set of the node $u$, $\alpha_{i}$ is the weight of the $i-th$ attribute, $v_i$ is the embedding of attribute $i$.

\subsection{Graph Embedding Networks for Encoding Graph Structures}\label{gnn:gen}
Graph embedding networks aim to represent graph structures of both the query graph and the provenance graph. There are two stages in the graph embedding networks: the node-level embedding stage and the graph-level embedding stage. We use two different graph embedding networks to encode the provenance graph $G_{p}$ and the query graph $G_{q}$ respectively. %重要The reason why we do not adopt two identical network structures is that as mentioned earlier the characteristics of two input graph $G_{q}$ and $G_{p}$ are very different.    
%The reason why we do not adopt the traditional Siamese network (which has two identical network branches and learns shared weights for them) is that as mentioned earlier the characteristics of two input graph $G_{q}$ and $G_{p}$ are very different.

\subsubsection{Stage 1: Node-level Embedding.}
\begin{figure*}
\centering
\includegraphics[width=1.0\textwidth]{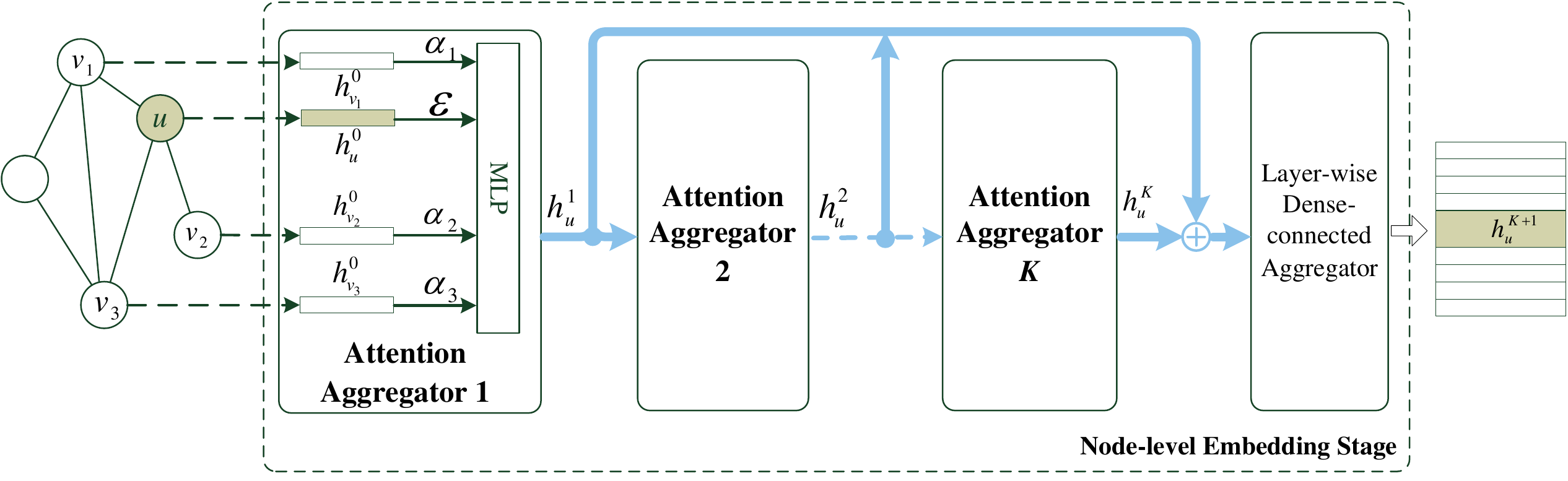}
\caption{Provenance graph node-level embedding network structure.} \label{fig4}
\end{figure*}

It is the node embedding method that leads to the difference between the query graph embedding network and the provenance graph embedding network. We adopt the existing graph convolutional network (GCN)~\cite{kipf2016semi} to embed the nodes in the query graph. Because the query graph is usually small and noise-free, which can be handled by GCN. To embed the nodes in the provenance graphs, we design a provenance graph embedding network structure (shown in Fig.~\ref{fig4}) which is capable of dealing with the redundant nodes while remaining the key information for matching the query graph. The provenance graph node embedding network is detailed as follows.

Firstly, we design a GNN layer, called $attention$ $aggregator$, which could aggregate information from neighbors for the target node. Note that, for a node $u$ in the provenance graph, different neighbors may have different importance for graph matching when incorporating their node features into node $u$. In particular, the redundant neighbors should be assigned a lower importance value to reduce their impact on $u's$ hidden representation $h_{u}$, while the nodes that can match the corresponding ones in the query graph should be assigned a higher importance value. For this purpose, we adopt another attention mechanism that can learn weights for neighbors of node $u$. Formally, the hidden representation $h_{u}$ outputted by the layer $k$ can be computed via a neural aggregation function that is achieved by

\begin{equation}
h^{k}_{u}=MLP(\epsilon ^{(k)}h^{k-1}_{u} + \sum\nolimits_{v \in N_{u}} \alpha_{v} h^{k-1}_{v}), \label{equ:node_emb}
\end{equation}

\noindent
where $h^{k}_{u}$ is the output of layer $k$ of the provenance graph embedding network, and it is the hidden representation of node $u$;  $\alpha_{v}$ is the attentional weight of node $v$ ($v \in N_{u}$, where $N_{u}$ is the set of node $u's$ neighbors).% We call Eq.~\ref{equ:node_emb} as attentional aggregator in Fig.~\ref{fig4}.%, and $\alpha_{v}$ determines how much the representation of previous layer that can be incorporated into the current layer.

To aggregate the information into $h_{u}$ from distant nodes, we then add more layers defined by Eq.~\ref{equ:node_emb}. The number of layers, $K$, means that the GNN can aggregate information from $u$'s $K-hops$ neighbors. However, simply adding layers may squash exponentially-growing information (including noise) into fixed-size vectors. To address this issue, we adopt the Layer-wise Dense-connected Aggregator which is proposed by~\cite{wang2019heterogeneous}. This strategy is formulated as follows:

\begin{equation}
h^{K+1}_{u}=MLP([h^{0}_{u};h^{1}_{u};\ldots h^{K}_{u}]),
\end{equation}

\noindent
where $[\cdot;\cdot]$ is the feature concatenation operation.

\subsubsection{Stage 2: Graph-level Embedding.}

%Now we obtain the hidden representation $h$ for each node in both the query graph and the provenance graphs. How to generate a low-dimensional embedding vector for a graph using node embeddings? In this work, we adapt the Global Context-Aware Attention strategy proposed in SimGNN~\cite{bai2019simgnn} to obtain the graph-level embedding. The idea is that the graph-level embedding is the weighted sum of node embeddings where a weight associated with a node is learned from the labeled graph pairs. Specifically, nodes that are similar to the global context should be assigned larger weights. Different from SimGNN, we normalize the weights into 1, because we do not want the graph size to affect the calculation of the matching score. Hence, we replace the sigmoid function in SimGNN with a softmax function $\sigma(\boldsymbol z)_{i}=\frac{e^{z_{i}}}{\sum \nolimits_{j=1}^{K} e^{z_{j}}}$. This graph-level embedding is formally represented by the following equation:
Now we obtain the node embedding $h$ for each node in both the query graph and the provenance graphs. How to generate a low-dimensional embedding vector for a graph using node embeddings? In this work, we adapt the Global Context-Aware Attention strategy proposed in SimGNN~\cite{bai2019simgnn} to obtain the graph-level embedding $h_{G}$. Intuitively, nodes that are similar to the global context will be assigned larger weights, which allows the corresponding node embeddings to contribute more to the graph-level embedding. Different from SimGNN, we normalize the weights into 1, because we do not want the graph size to affect the calculation of the matching score. Hence, we replace the sigmoid function in SimGNN with a softmax function $\sigma(\boldsymbol z)_{i}=\frac{e^{z_{i}}}{\sum \nolimits_{j=1}^{K} e^{z_{j}}}$. This graph-level embedding is formally represented by the following equation:

\begin{equation}
\begin{aligned}
h_{G}&=\sum\nolimits_{u=1}^{N} \sigma(h_{u}c)h_{u}\\
&= \sum\nolimits_{u=1}^{N} \sigma(h_{u}\tanh((\frac{1}{N} \sum\nolimits_{m=1}^{N} h_{m})W))h_{u},
\end{aligned}
\end{equation}

\noindent
where $N$ is the number of nodes in a graph, $\tanh(\cdot)$ is a activation function, $W$ is the trainable parameters.

%\subsection{Put It All Together}\label{gnn:archi}
\subsection{GNN-based Architecture for Graph Pattern Matching}\label{gnn:archi}
\begin{figure*}
\centering
\includegraphics[width=1.0\textwidth]{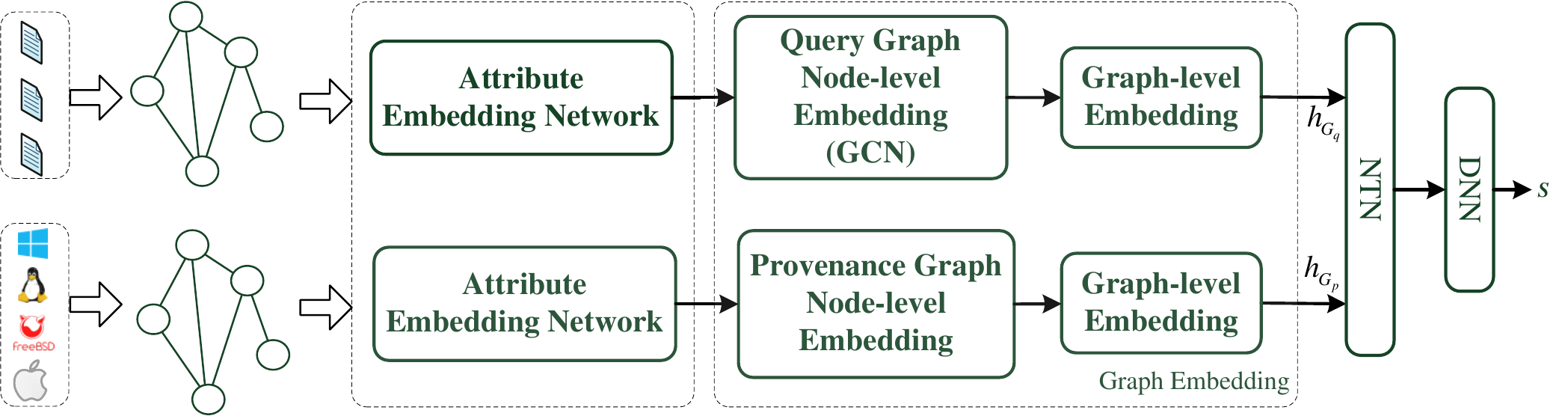}
\caption{The framework of DeepHunter's graph pattern matching model.} \label{fig3}
%\caption{GNN-based graph pattern matching model.} \label{fig3}
\end{figure*}

Based on the attribute embedding network and the graph embedding network, DeepHunter's graph pattern matching model could learn robust graph patterns.
The framework of DeepHunter's graph pattern matching model is shown in Fig.~\ref{fig3}. It consists of two branches. The upper branch of Fig.~\ref{fig3} deals with CTI information, and the lower one is for provenance data. At the beginning of each branch, the query graph and the provenance graph are constructed. Then both of them are fed into our GNN-based models.

Given the output of two branches, $h_{G_{q}}$ and $h_{G_{p}}$, many existing graph matching models adopt the Siamese architecture~\cite{bromley1994signature} to learn the relation between them. However, the Siamese architecture that directly computes the inner product of $h_{G_{q}}$ and $h_{G_{p}}$ is too simple to model the complex relation. Instead, we employ Neural Tensor Network (NTN), which is a powerful relation learning network, to replace the inner product operation. We compare NTN and the traditional Siamese architecture in section~\ref{eva:comparewithotherGM}.

After the NTN layer, we connect multi-layer dense neural networks (DNNs) and output the graph matching score $s$. At last, to compute the loss, we compare $s$ against the ground-truth label using the following mean squared error loss function:
\begin{equation}
\mathcal{L}=\sum_{(G_{p_{i}},G_{q_{i}})\in \mathcal{D}} (\hat{s} - s(h_{G_{p_{i}}}, h_{G_{q_{i}}}))^{2},
\end{equation}
where $ D=\{(G_{p_{1}},G_{q_{1}}),(G_{p_{2}},G_{q_{2}}),...\} $ is the training dataset.

We train the proposed model in an end-to-end way. We leverage stochastic gradient descent to estimate parameters. After a number of training epochs, the loss value will be small and stable, the accuracy of validation data will be high, which demonstrates that the model is trained well.

%% file: implement-4.tex
\section{Implementation}\label{datapreparation}

\subsection{Provenance Graph Reduction} \label{imp:preprocessProvenance}

In practice, due to hosting long-term system logs is prohibitively expensive, analysts attempt to reduce the provenance graph and yet preserve the quality of threat hunting~\cite{hassan2020tactical}.
In this work, we prune the provenance graph as follows.

% /////////////////////////////////////////////////////////////
First, we leverage the MITRE ATT\&CK TTPs and the IOCs to generate suspicious events.
Specifically, DeepHunter uses the EDR tool (i.e., BLUESPAWN~\cite{bluespawn}), which provides matching rules to detect MITRE ATT\&CK TTPs. 
Besides, DeepHunter also matches the IOCs (extracted from threat intelligence, such as APT reports) using regular expressions. 
The events identified by both the EDR tool and the IOC matching are regarded as suspicious events.

% /////////////////////////////////////////////////////////////
We then propose the provenance graph reduction algorithm (Algorithm~\ref{Alg_1}), which could prune the provenance graphs based on the suspicious events. 
Inspired by Poirot~\cite{milajerdi2019poirot}, we select $ seed $ $ nodes $ from the nodes that match the IOCs. 
For example, suppose IOC $\alpha$ has $x$ matched nodes in provenance graphs, IOC $\beta$ has $ y $ matched nodes, and IOC $\gamma$ has $ z $ matched nodes. If $ z=min\{x,y,z\} $, then these $ z $ nodes are $ seed$ $nodes $.
%(When an IOC has the smallest number of alert nodes among all the indicators, these matching nodes are $seed$ $nodes$.) 
We start from a $seed$ $node$ and execute $adaptiveBFS$ searching on the provenance graphs. 
A suspicious subgraph generated by the graph reduction algorithm could cover all IOCs' alerts.
%The proposed algorithm is shown in Algorithm~\ref{alg:Framwork}.

\begin{algorithm}[!t]
\SetAlgoLined
\DontPrintSemicolon
\caption{ Provenance Graph Reduction Algorithm }\label{Alg_1}
\label{alg:Framwork}
\KwIn{Provenance Graphs: $G_{p}$, Indicators Set: $I$, Matched Nodes Set: $P$;}
\KwOut{Suspicious Subgraphs: $SuspGraphs$}

    \SetKwFunction{FMain}{ExpandSearch}
    \SetKwProg{Fn}{Function}{:}{}
    \Fn{\FMain{$SeedNodes$, $Susp$}}{
    	\ForEach{$ node \in SeedNodes $}
    	{
			$start\_node \gets node$;
			$subgraph \gets adaptiveBFS(start\_node ,P)$;
			$Susp \gets ComposeGraph(Susp, subgraph)$;
			
			\eIf{$Susp$ contains all indicators in $I$}
            {
            	Add $Susp$ to $SuspGraphs$           
            }
            {
            	$remain\_nodes$ $\gets$  $seed$ $nodes$ from $P$ that are not matched with any indicators in $I$;
            	$ExpandSearch(remain\_nodes, Susp)$;
            }
    	}
    	
%        $ S^{*}  \longleftarrow F $;    
%
%        \ForEach{$ F \in NPs $}
%        {\eIf{$ f_i = Negated $}
%            {$ N \longleftarrow f_i;$}
%            {$ S \longleftarrow f_i;$}
%
%        }
        \textbf{return} $SuspGraphs$
}
%\textbf{End Function}
\end{algorithm}

% /////////////////////////////////////////////////////////////
The $adaptiveBFS$ is an adapted Breadth-First Search (BFS) algorithm. 
Specifically, during BFS on the provenance graph, only the nodes related to suspicious events and the process node could be visited. 
%Algorithm~\ref{alg:Framwork} is very efficient. 
%We evaluate its efficiency in section~\ref{efficiency}.

% /////////////////////////////////////////////////////////////
Obviously, the suspicious subgraphs generated by our provenance graph reduction algorithm contain lots of false positives (the threat alert fatigue problem). 
Therefore, it is still necessary for analysts to use our graph pattern matching model (Section~\ref{gnn}) to calculate the matching score. 
%In the evaluation (section~\ref{eva:preprocessProvenance}), we apply our reduction method for different attack scenarios. 
%The results demonstrate that Algorithm~\ref{alg:Framwork} is effective and necessary for threat hunting.

\subsection{Training Data Generation} \label{Generation of training graph pairs and learning parameters}
Training DeepHunter requires a large number of positive samples $(G_{p_i}, G_{q_i})$ ($\mathcal{M}(G_{p_i}, G_{q_i})=1$) and negative samples $(G_{p_i}, G_{q_i})$ ($\mathcal{M}(G_{p_i}, G_{q_i})=-1$). The query graph can be considered as a summarization of its corresponding provenance graph. Therefore, we use the graph summarization techniques to generate the matched query graph $G_q$ for each provenance graph $G_p$. We also add random noise to improve the robustness. We detail the training data generation method as follows.

Firstly, we extract a subgraph as $G_{p_{i}}$ from the provenance graphs. Specifically, we start from a process node and use DFS on the provenance graph. We limit the length of the paths, which is less than 4.
Then we refine the $G_{p_{i}}$ using two graph summarization rules as follows: 

\begin{itemize}
	\item	Merge process nodes that have the same process name;
	\item	Remove duplicate paths. If two paths are duplicates (i.e., two sequences of node name are equal), only one is reserved.
\end{itemize}

Then we add noise to $G_{p_{i}}$ by:
\begin{itemize}
     \item	randomly dropping edges or object nodes on $G_{p_{i}}$;
     \item	randomly removing one or more attributes of a node.
\end{itemize}

After the above two steps, we generate a $G_{q_{i}}$ for the $G_{p_{i}}$. So $ \left( G_{q_{i}}, G_{p_{i}}\right) $ is a positive sample for training.

%////////////////////////////////////////////////////////////////////////////
At last, we construct the negative sample $ ( G_{q_{j}}, G_{p_{i}}) $ by randomly combining $G_{p_{i}}$ and $G_{q_{j}}$, where $ i\neq j $.
By doing this, we simulate the situation where most of the node attribute information and the main graph structure of $G_{p_{i}}$ is preserved in $G_{q_{i}}$. %The performances of the graph pattern matching models (learning-based) shown in Table~\ref{Tabel6} demonstrate that our training dataset generation method is effective.

%% file: evaluation-5.tex
\section{EVALUATION}\label{eva}

\subsection{Attack Scenarios and Experimental Setup}

To evaluate the efficacy of DeepHunter, we utilize provenance data which contain 5 APT attack scenarios, including 3 real-life APTs(DARPA TC engagement 3) and 2 synthetic APTs. For each of the attack scenarios, the corresponding query graph is also provided. To simulate real-world threat hunting, the query graphs we used in the evaluation are either generated by the third-party or constructed based on the public APT reports. The description of APT scenarios and the source of corresponding query graphs are shown in Table~\ref{attackscenario}.%In order to simulate the real-world threat hunting, the query graphs we used in the evaluation are either defined by the third-party or constructed based on the public APT reports (For DARPA dataset, we use the ground truth reports provided by DARPA TC engagement 3).

\begin{table*}[t!]
\caption{APT attack scenarios description and the source of query graphs.}
\renewcommand{\arraystretch}{1.1}
%\resizebox{\textwidth}{!}{
\begin{tabular}{c|p{7.5cm}|p{2.2cm}}   
%\begin{tabular}{c|c|p{2.8cm}|c}   % p{x com} x��ʾ����xcm�Զ�����
\hline
\textbf{Scenario}    & \textbf{Short Description}   & \textbf{Query Graph Source}  \\ \hline \hline
        Q1+CADETS           & A Nginx server was exploited and a malicious file was downloaded and executed. The attacker tried to inject into sshd process, but failed.  & DARPA TC 3 reports\\ \cline{1-3}
      
        Q2+TRACE            & The Firefox process was exploited and established a connection to the attacker's operator console. The attacker downloaded and executed a malicious file.
& DARPA TC 3 reports \\ \cline{1-3}

        Q3+TRACE            & A Firefox extension (a password manager) was exploited. A malicious file was downloaded and executed to  connect out to the C\&C server. 
& DARPA TC 3 reports \\ 
\cline{1-3}

        Q4+ETW              & Detailed in section~\ref{bme:motivatingexample}.
& Fig.~\ref{fig1} with persistence \uppercase\expandafter{\romannumeral1}\\ 
\cline{1-3}

        Q5+ETW              & The attack mutation of scenario Q4. 
& Fig.~\ref{fig1} with persistence \uppercase\expandafter{\romannumeral2}\\ \hline \hline
\end{tabular}

\label{attackscenario}
\end{table*}

%重要The datasets in our paper are used to evaluate the accuracy and robustness of DeepHunter, especially the latter. For the evaluation of robustness, we perform a comparison experiment in Section~\ref{eva:robustness}.  In particular, we compare DeepHunter to Poirot, a state-of-the-art APT threat hunting approach. The results demonstrate that DeepHunter is resistant to the inconsistency between the query graph and the provenance graph caused by attacks or the provenance systems. %Additionally, we also analyze how the inconsistency impairs the performance of Poirot. %, or the inaccurate threat intelligences

\subsubsection{Inconsistency Scores.} 
Before evaluating robustness, we define three inconsistency scores to quantify the degree of the inconsistency between the query graph and the corresponding provenance graph. Specifically, we compute graph edit distance (GED) and the number of $missing$ $nodes$ and $missing$ $paths$. GED measures the cost that transforms $G_q$ into $G_p$. We adopt a graph matching toolkit~\cite{riesen2013novel,graphmatchingtookit-github} to calculate GED and normalize~\cite{qureshi2007graph} the GED scores for different graph sizes. The $missing$ $node$ of the query graph is the node that we cannot find its alignments in provenance graphs. The $missing$ $path$ means that for an edge from node $i$ to $j$ in the query graph, there is no path from the nodes aligned to $i$ to the nodes aligned to $j$ in provenance graphs. Table~\ref{Tabel2} shows the inconsistency scores of the scenarios in Table~\ref{attackscenario}. We can see from Table~\ref{Tabel2} that the chosen scenarios contain different degrees of inconsistency.
\begin{table}[]
\caption{Inconsistency scores of different scenarios. The values in parentheses on the second and third columns are the number of missing nodes and paths, respectively.}
\center
\renewcommand{\arraystretch}{1.1}
\begin{tabular}{p{2.5cm}<{\centering}|p{3.1cm}<{\centering}|p{3.1cm}<{\centering}|p{1.5cm}<{\centering}}   % p{x com} x��ʾ����xcm�Զ�����
%\begin{tabular}{c|c|p{2.8cm}|c}   % p{x com} x��ʾ����xcm�Զ�����
\hline
\textbf{Scenario}  & \textbf{Missing Nodes (\%)}        & \textbf{Missing Paths (\%)}                             &        \textbf{GED}        \\ \hline \hline 
        Q1+CADETS  &              0                    &                0                          		    &          0.192   \\ \cline{1-4}
        Q2+TRACE    &               0                  &               6.25\%(1)                            &          0.303 \\ \cline{1-4}
        Q3+TRACE    &               0                  &               16\%(4)                              &          0.504 \\ \cline{1-4}
        Q4+ETW        &               3.8\%(1)         &               4\%(1)                               &         0.454                     \\ \cline{1-4}
        Q5+ETW        &           21.4\%(6)            &               20\%(7)                              &           0.557                    \\ \hline \hline
\end{tabular}
\label{Tabel2}
\end{table}

%\begin{figure}[htbp]
%\centering
%
%\subfigure[]
%{
%	\begin{minipage}[t]{0.21\textwidth}
%	\centering          %��ͼ����
%	\includegraphics[width=\textwidth]{figx2_left.pdf}   %��pic.jpg��0.5����С����
%    \label{figx2_left}
%	\end{minipage}
%}
%%\hspace{0.1in}
%\subfigure[]
%{
%	\begin{minipage}[t]{0.21\textwidth}
%	\centering      %��ͼ����
%	\includegraphics[width=\textwidth,height=3.5cm]{figx2_right.pdf}   %��pic.jpg��0.5����С����
%    \label{figx2_right}
%	\end{minipage}
%}
%
%%\hspace{0.1in}
%
%\caption{Scenario Q1: left is the query graph and right is the provenance graph of the attack in scenario Q1.} %  %��ͼ����
%\label{figx2}  %ͼƬ���ñ���
%\end{figure}

\subsubsection{Experimental Setup.} 
The provenance data from DARPA are collected by two provenance systems: CADETS~\cite{cadetswebpage} and TRACE~\cite{tracewebpage}. Besides, we synthesized attacks in scenarios Q4+ETW and Q5+ETW on Windows 7 32 bit systems. The provenance data of Q4+ETW and Q5+ETW, including benign system activities and attack behaviors, were collected by our ETW-based provenance system. 

We employed the gensim~\cite{rehurek_lrec} Python library to obtain the attribute embeddings $ v_{i} $ (detailed in section~\ref{gnn:aen}). We implemented the proposed graph neural network model using PyTorch~\cite{NEURIPS2019_9015}. We trained whole neural network-based models using 2 Nvidia Tesla P4 GPU. Other experiments (e.g., provenance graph construction, graph reduction, etc.) are conducted on a server with two Intel Xeon E5-2630 v3 CPUs and 128 GB memory running CentOS system. %We generate a dataset for each provenance system, including training data and test graphs (suspicious subgraphs generated by our graph reduction method). 

\subsubsection{Datasets.}
We generated a dataset for each provenance system and named the dataset after the provenance system. 
We used the graph reduction method illustrated in section~\ref{imp:preprocessProvenance} to prune the provenance graph. The generated subgraphs (i.e., test graphs) were manually labeled based on the corresponding reports' timestamp. We also generated training graph pairs using the method detailed in section~\ref{Generation of training graph pairs and learning parameters}. The characteristics of our datasets are shown in Table~\ref{eva_table:dataset}.

\begin{table}[]
\center
\caption{The characteristics of graph datasets used in our evaluation.}
%\resizebox{120mm}{!}{
\renewcommand{\arraystretch}{1.1}
\begin{tabular}{c|c|p{13.5mm}<{\centering}|c|c}   % p{x com} x��ʾ����xcm�Զ�����
%\begin{tabular}{c|c|p{2.8cm}|c}   % p{x com} x��ʾ����xcm�Զ�����
\hline
\multirow{2}{*}{\textbf{Dataset}}    &  \multirow{2}{*}{\textbf{Raw Graph Size}}  & \multicolumn{2}{c|}{\textbf{\# of Test Graphs}} & \multirow{2}{*}{\textbf{\# of Training Graphs}}  \cr \cline{3-4}
& & Benign & Attack & 
\\ \hline \hline
        CADETS &       904 MB        &           10     &   1      &      150,000 \\ \cline{1-5}
        TRACE  &       22.5GB        &           9      &   6      &      150,000 \\ \cline{1-5}
        ETW    &       40GB          &           105    &   10     &      300,000 \\ \hline \hline
\end{tabular}%}
\label{eva_table:dataset}
\end{table}

\subsection{Robustness}\label{eva:robustness}
We evaluate the impact of inconsistency on DeepHunter and the state-of-the-art Poirot. We also analyze why Poirot fails in scenario Q5+ETW which contains disconnected attack provenance graphs. %aim is to see whether DeepHunter can tolerate the inconsistency (quantified in Table~\ref{Tabel2}) between the query graph and the provenance graph. We also evaluate the impact of inconsistencies on Poirot for comparison.

\subsubsection{State-of-the-art Poirot.} 
Poirot is a heuristic graph pattern matching algorithm that can compute the graph alignment score between the query graph and the provenance graph. Poirot searches for aligned nodes in the provenance graphs according to the $information$ $flows$ in the query graph. During the search, Poirot omits the paths that are impossible to be adopted by attackers. %In other words, each visited path $p$ in the attack provenance graphs must guarantee that the number of $launcher$ nodes is less than a specific value (which is set to 3 in Poirot). %The launcher node represents an entry point process that an attacker is willing to exploit independently.%It searches suspicious nodes on the provenance graphs alongside paths that consistent with the $information$ $flows$ in the query graph.

\subsubsection{Experimental Results.} 
\begin{figure}[t]
%\vspace{-0.4cm}
\centering
\includegraphics[scale=0.45]{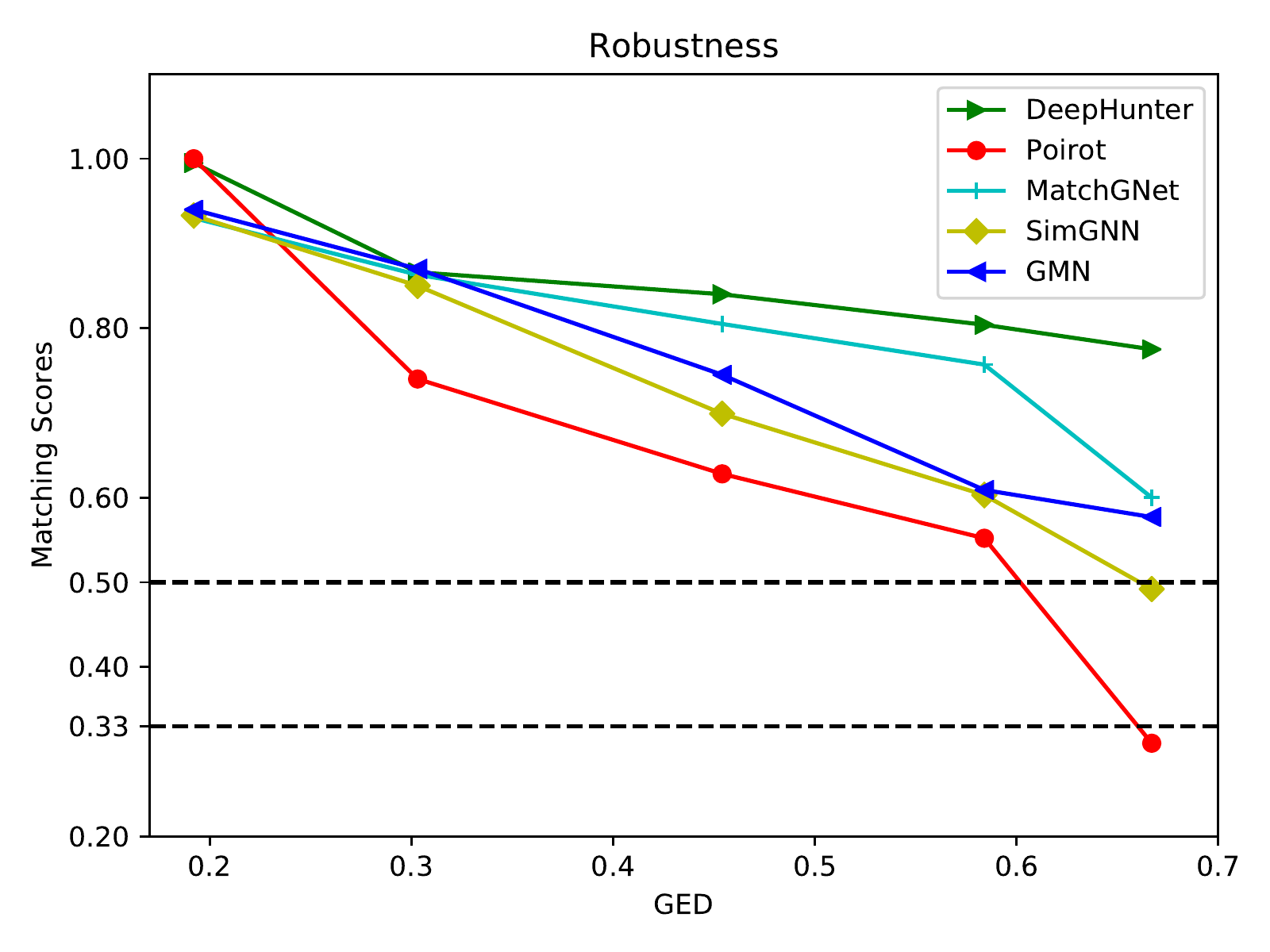}
\vspace{-0.4cm}
\caption{Axis x: graph edit distance between $G_{q}$ and $G_{p}$ of each scenario in Table~\ref{attackscenario}; Axis y: matching scores between $G_{q}$ and $G_{p}$ of each scenario.} %The result comes from Table~\ref{Tabel5}.}  
\label{figx4}
\end{figure}

We compare DeepHunter with Poirot using all scenarios in Table~\ref{attackscenario}. The matching scores calculated by Poirot, DeepHunter, and other GNN-based graph matching models are shown in Fig.~\ref{figx4}. We can see that all matching scores calculated by DeepHunter are greater than the threshold (which is 0.5). This result shows that the accuracy of DeepHunter can be guaranteed in scenarios where there exist various degrees of inconsistency.
%when there is a certain degree of inconsistency between the qurery graph and the provenance graph. 

Moreover, as the degree of the inconsistency increases, all matching scores decrease. But the curve of DeepHunter is more stable. On the contrary, the curve of Poirot drops faster than the GNN-based graph matching models. Even worse, the matching score calculated by Poirot is less than its threshold in the most inconsistent scenario Q5+ETW , which means that Poirot fails to identify this attack.

Additionally, we detail the false positive results of DeepHunter and Poirot in Table~\ref{fp_results}. The results can demonstrate that the high robustness of DeepHunter is not built upon false positives.

\begin{table*}[]
\center
\caption{False positive results of DeepHunter and Poirot.}
%\resizebox{122mm}{!}{
\renewcommand{\arraystretch}{1.0}
\begin{tabular}{c|c|c|c}   % p{x com} x��ʾ����xcm�Զ�����
%\begin{tabular}{c|c|p{2.8cm}|c}   % p{x com} x��ʾ����xcm�Զ�����
\hline
\textbf{Dataset}        & \textbf{CADETS}  & \textbf{TRACE} & \textbf{ETW}  \\ 
\hline  \hline 
\# of Test Graphs       &           11     &        15      &      115      \\ \cline{1-4}
\# of FPs(DeepHunter)   &           0      &         0      &      1        \\ \cline{1-4}
\# of FPs(Poirot)       &           0      &         1      &      2        \\ \hline \hline
\end{tabular}%}
\label{fp_results}
\end{table*}

\subsubsection{Why Poirot Fail?} 
When searching on the disconnected attack provenance graphs in Q5+ETW, the paths which start from nodes belonging to the EternalBlue exploitation stage to nodes belonging to the cryptocurrency mining stage can not be found. So the $graph$ $alignment$ $score$ computed using the Equation (2) in Poirot becomes smaller (The $influence$ $scores$ of the missing paths are all equal to 0. And the denominator of Equation (2), $|F(G_{q})|$, which is the number of $flows$ in the query graph, remains unchanged). As a result, this type of inconsistency in Q5+ETW leads to the invalidation of Poirot. On the contrary, DeepHunter does not rely on complete connectivity remained in the provenance graph. As long as most node attribute information and the main graph structure information between the query graph and the traceability graph are matched, DeepHunter can recognize that the two graphs represent the same attack behavior. Therefore, DeepHunter has a more robust cyber threat hunting ability. %to this type of inconsistency, because DeepHunter does not dependent on the complete attack paths remained in the provenance graph. As long as the query graph is mainly summarizing the provenance graph, DeepHunter could achieve good performance.%(需要大修图2，使细节更多) %, due to the disconnection between two subgraphs. %For the same reason, we believe that the performance of other path-based approaches could also be impaired under the similar circumstance. 

\subsection{Comparison with Other Graph Matching Models}\label{eva:comparewithotherGM}
We compare DeepHunter with a non-learning graph matching approach and other GNN-based graph matching models. Note that these GNN-based models are not specifically designed for threat hunting. 
%Note that these approaches (Table~\ref{Tabel6}) use suspicious subgraphs generated by our graph reduction method to calculate the matching score. 
%So the test set is composed of graph pairs, each of which consists of the given query graph and the suspicious subgraph. (The number of suspicious subgraphs for each scenario is shown in the second column of Table~\ref{Tabel4}.) 
We evaluate all the graph matching approaches using the AUC value, since it is a strict metric. If a small mistake is made, the error would be obvious. %the are used to match the query graph and the preprocessed provenance graphs shown in Table~\ref{Tabel4}. %So the number of suspicious graphs (shown in the second column of Table~\ref{Tabel4}) is the number of test graphs %  is the number of test graphs for the evaluation. for the evaluation. %Note that the approaches in this section calculate the matching score between the query graph and the suspicious subgraph generated by our graph reduction method. So the test set is composed of these suspicious subgraphs (shown in the second column of Table~\ref{Tabel4}).

\subsubsection{DeepHunter vs. Non-learning Approach.} 
We compare DeepHunter with the Weisfeiler Lehman (WL) kernel, a non-learning method for calculating the graph similarity. 
%WL kernel also creates an embedding for each graph, then computes the inner product of the two graph embeddings as the matching score. 
We set the number of iteration of the WL kernel from 1 to 10 and put the best results in Table~\ref{Tabel6}. The result of the WL kernel is not desirable because it is designed for graph isomorphism testing. 
In contrast, the graph matching in a threat hunting task is more similar to determining whether a query graph can be regarded as an abstraction of the provenance graph.%the provenance graph can be summarized to the query graph. %We can see that by learning on the various distributions of graph pairs (Fig. x3), DeepHunter is able to perform better than traditional graph kernel method WL kernel which is a state-of-the-art graph kernel. %tried the parameter $h$ (the iteration time) of WL kernel from 1 to 10 and put the best results in Table~\ref{Tabel6}. 

\subsubsection{DeepHunter vs. GNN-based Graph Matching Models.} 
We compare DeepHunter's graph matching model with other GNN-based graph matching networks: MatchGNet~\cite{wang2019heterogeneous}, SimGNN~\cite{bai2019simgnn} and GMN~\cite{li2019graph}. MatchGNet proposed a Hierarchical Attentional Graph Neural Encoder (HAGNE) which could embed the provenance graph. Given the graph-level embeddings, MatchGNet employs the Siamese network to learn the similarity metric. We believe that the Siamese network is not enough to learn the complex relationship between the two graphs. Hence, we substitute the Siamese network of MatchGNet with the NTN layer. 
We call the modified model MatchGNet-NTN. By the comparison between DeepHunter and MatchGNet-NTN, the effectiveness of our graph embedding networks can be testified. 
As can be seen in Table~\ref{Tabel6}, the performance of DeepHunter outperforms MatchGNet and MatchGNet-NTN. 

We also evaluate the effectiveness of the attribute embedding network. Instead of the attribute embedding network, we directly use the one-hot encoding of attributes as the node's input feature. We call this model DeepHunter-wo-AEN. Table~\ref{Tabel6} shows that DeepHunter-wo-AEN is inferior to DeepHunter, which demonstrates the attribute embedding network is necessary for our graph matching task.

%We also compare DeepHunter with other GNN-based graph matching networks: MatchGNet~\cite{wang2019heterogeneous}, SimGNN~\cite{bai2019simgnn} and GMN~\cite{li2019graph}. MatchGNet proposed a Hierarchical Attentional Graph Neural Encoder (HAGNE) which can be uesd to embed the provenance graph. But the architecture of MatchGNet can not be used directly for the threat hunting task, because the output of HAGNE is the node embedding, not the graph-level embedding. We add the stage 2: graph-level embedding of our Graph Embedding Network after the HAGNE. Given the graph-level embeddings, MatchGNet employs the Siamese network to learn the similarity metric. We believe that the Siamese network is not enough to learn the complex relationship between two graphs. Hence, we substitute the Siamese network with the NTN layer. We name the modified model as MatchGNet-NTN. By the comparison between DeepHunter and MatchGNet-NTN, the effectiveness of our Graph Embedding Networks can be testified. As can be seen in Table~\ref{Tabel6}, the performance of DeepHunter outperforms MatchGNet and MatchGNet-NTN. We also evaluate the usefulness of our Attribute Embedding Network. We drop the Attribute Embedding Network, and directly use the one-hot encoding of attributes as the input feature of nodes. We name this model as DeepHunter-wo-AEN. The result in Table~\ref{Tabel6} shows that AEN is necessary for our graph match task.

At last, we evaluate the other two graph matching networks: SimGNN and GMN. Like DeepHunter, SimGNN also leverages GNN to represent input graphs and then utilizes NTN to learn the similarity between two graph-level embeddings. But the graph neural networks in SimGNN are not specifically designed for representing the provenance graphs. Besides, SimGNN believes that if there is a difference in the size of the two input graphs, then the two graphs are not similar. GMN takes into account the node correlation across graphs to model the relation. The results of SimGNN and GMN are shown in Table~\ref{Tabel6}. We can see that the performance of DeepHunter is superior to both SimGNN and GMN. %Specifically, SimGNN also utilizes NTN to model the similarity between embeddings of two graphs. While GMN take into account the node correlation across graphs. The results of these two GNN-based model are shown in Table~\ref{Tabel6}. We can see that the performance of DeepHunter is better than both of these graph matching networks. 

\begin{table*}[]
\center
\caption{AUC values of graph matching models on three datasets.}
%\resizebox{122mm}{!}{
\renewcommand{\arraystretch}{1.1}
\begin{tabular}{c|c|c|c}   % p{x com} x��ʾ����xcm�Զ�����
%\begin{tabular}{c|c|p{2.8cm}|c}   % p{x com} x��ʾ����xcm�Զ�����
\hline
\textbf{Dataset}        & \textbf{CADETS}  & \textbf{TRACE} & \textbf{ETW}  \\ 
\hline 
DeepHunter     &    1             & \textbf{0.951} & \textbf{0.916}  \\ \cline{1-4}
MatchGNet~\cite{wang2019heterogeneous} &1  &         0.880  &         0.805   \\ \cline{1-4}
MatchGNet-NTN           &    1             &         0.932  &         0.844   \\ \cline{1-4}
MatchGNet-wo-AEN        &    1             &         0.891  &         0.820   \\ \cline{1-4}
SimGNN~\cite{bai2019simgnn}& 1             &         0.906  &         0.805   \\ \cline{1-4}
GMN~\cite{pmlr-v97-li19d}&   1             &         0.846  &         0.830   \\ \cline{1-4}
WL kernel                &   1             &         0.492  &         0.301   \\ \hline 

\end{tabular}%}
\label{Tabel6}
\end{table*}